\documentclass[12pt]{article}

\usepackage[utf8]{inputenc}
\usepackage{amsmath, amssymb}
\usepackage{float}
\usepackage{graphicx}
\usepackage{authblk}
\usepackage{hyperref}
\usepackage{cleveref}
\usepackage{geometry}
\usepackage[table]{xcolor}
\usepackage{newunicodechar}
\usepackage{tcolorbox}

\usepackage{chngcntr}
\usepackage{newfloat}

\DeclareFloatingEnvironment[
  fileext=los,
  listname={List of Supplementary Tables},
  name=Supplementary Table,
  placement=tbhp,
]{supplementarytable}

\DeclareFloatingEnvironment[
  fileext=lof,
  listname={List of Supplementary Figures},
  name=Supplementary Figure,
  placement=tbhp,
]{supplementaryfigure}

\crefname{supplementarytable}{Supplementary Table}{Supplementary Tables}
\Crefname{supplementarytable}{Supplementary Table}{Supplementary Tables}
\crefname{supplementaryfigure}{Supplementary Figure}{Supplementary Figures}
\Crefname{supplementaryfigure}{Supplementary Figure}{Supplementary Figures}
\crefname{supplementarytext}{Supplementary Text}{Supplementary Texts}
\Crefname{supplementarytext}{Supplementary Text}{Supplementary Texts}

\newunicodechar{₂}{$_2$}

\title{aLLoyM: A large language model for alloy phase diagram prediction}

\author[1]{Yuna Oikawa}
\author[2]{Guillaume Deffrennes}
\author[3]{Taichi Abe}
\author[4,1]{Ryo Tamura\thanks{Corresponding author. Email: tamura.ryo@nims.go.jp, }}
\author[1,4,5]{Koji Tsuda\thanks{Corresponding author. Email: tsuda@k.u-tokyo.ac.jp}}

\affil[1]{Graduate School of Frontier Sciences, The University of Tokyo, 5-1-5 Kashiwa-no-ha, Kashiwa, Chiba, 277-8561, Japan}
\affil[2]{University Grenoble Alpes, CNRS, Grenoble INP, SIMaP, Grenoble, F-38000, France}
\affil[3]{Research Center for Structural Materials, National Institute for Materials Science, 1-2-1 Sengen, Tsukuba, Ibaraki, 305-0047, Japan}
\affil[4]{Center for Basic Research on Materials, National Institute for Materials Science, 1-1 Namiki, Tsukuba, Ibaraki, 305-0044, Japan}
\affil[5]{RIKEN Center for Advanced Intelligence Project, 1-4-1 Nihonbashi, Chuo-ku, Tokyo, 103-0027, Japan}

\date{}

\begin{document}

\maketitle

\newpage

\begin{abstract}

Large Language Models (LLMs) are general-purpose tools with wide-ranging applications, including in materials science. In this work, we introduce aLLoyM, a fine-tuned LLM specifically trained on alloy compositions, temperatures, and their corresponding phase information. To develop aLLoyM, we curated question–and–answer (Q\&A) pairs for binary and ternary phase diagrams using the open-source Computational Phase Diagram Database (CPDDB) and assessments based on CALPHAD (CALculation of PHAse Diagrams).
We fine-tuned Mistral, an open-source pre-trained LLM, for two distinct Q\&A formats: multiple-choice and short-answer. Benchmark evaluations demonstrate that fine-tuning substantially enhances performance on multiple-choice phase diagram questions. Moreover, the short-answer model of aLLoyM exhibits the ability to generate novel phase diagrams from its components alone, underscoring its potential to accelerate the discovery of previously unexplored materials systems.
To promote further research and adoption, we have publicly released the short-answer fine-tuned version of aLLoyM, along with the complete benchmarking Q\&A dataset, on Hugging Face.

\end{abstract}

\newpage

\section*{Introduction}

Phase diagrams serve as fundamental roadmaps in materials science, providing critical insights into material behavior across varying thermodynamic conditions. The ability to accurately predict and interpret these diagrams represents a cornerstone of efficient materials design, with experienced practitioners often relying on accumulated expertise to anticipate phase relationships. While large experimental databases\cite{Schlesinger1983-lt,Massalski1990-kj,Villars1995-zj,Okamoto2010-wj} and computational repositories\cite{CPDDB,Jung2020,Hallstedt2023} have established valuable reference collections, the experimental determination of phase diagrams remains resource-intensive and prohibitively time-consuming for comprehensive materials exploration.

Recent advances in machine learning methodologies have demonstrated promising capabilities for phase diagram prediction, with conventional approaches including neural networks, support vector machines, random forests, and label propagation algorithms showing measurable success\cite{Terayama2019,Aghaaminiha2020,Dai2020,Lund2022,Zipoli2022,Tamura2022,Deffrennes2022,Tamura2024}. Concurrently, the emergence of large language models (LLMs) such as GPT-4, LLaMA, and Mistral has opened novel avenues for materials science applications\cite{Jablonka2023,Liu2023,Lei2024,Deb2024,Jiang2025}. Unlike specialized machine learning models that operate on isolated datasets, LLMs represent general-purpose architectures capable of leveraging broader scientific knowledge such as thermodynamic principles and elementary properties encoded during pre-training, into phase diagram predictions. Preliminary investigations have explored LLM applications in phase diagram analysis, including system-specific training on Mg-Al-Zn data\cite{Yan2024} and experimental diagram annotation\cite{Zha2024}, suggesting substantial potential for phase diagram analysis.

\begin{figure}[H]
    \centering \includegraphics[width=0.8\linewidth]{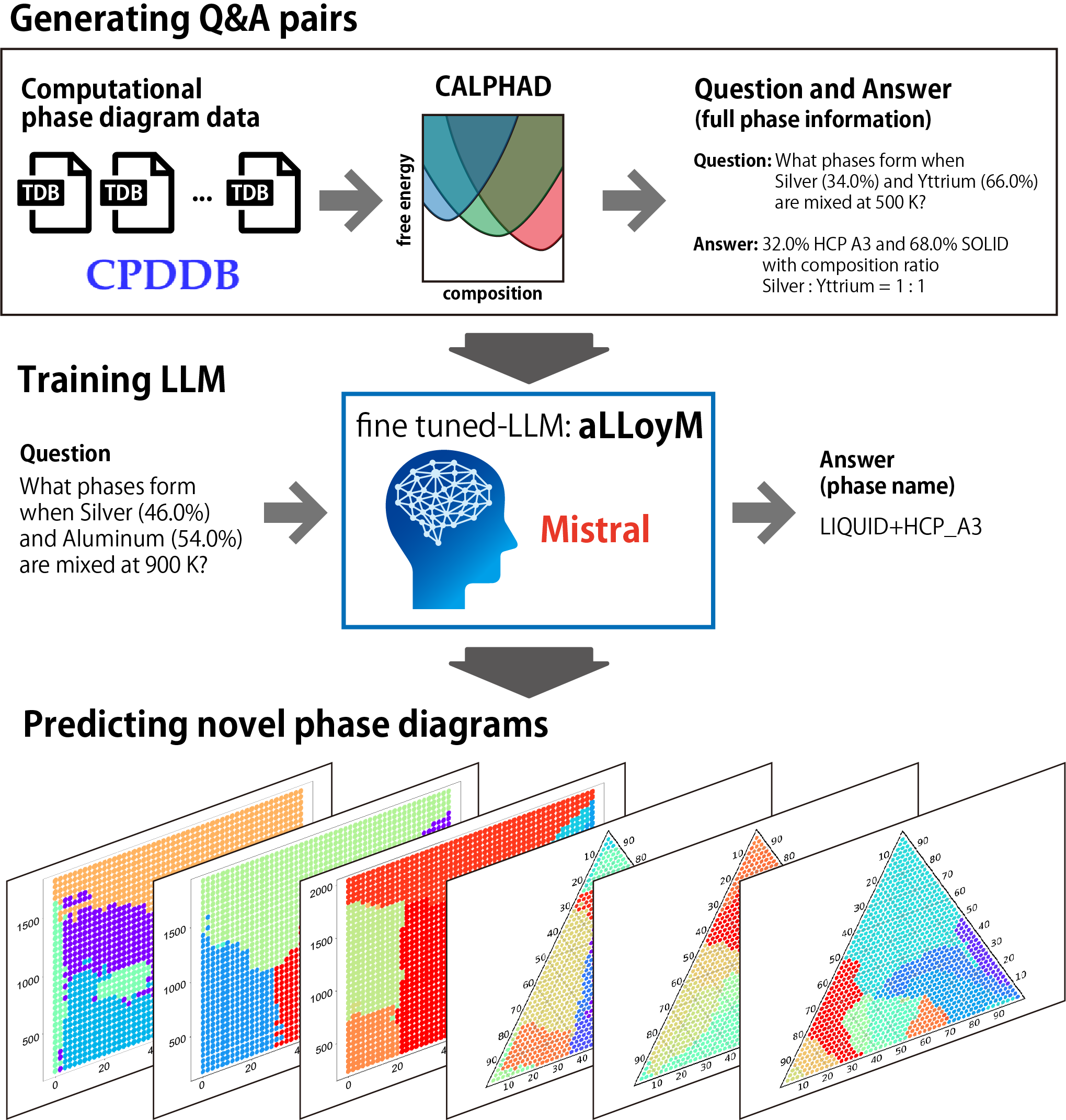}
    \caption{Schematic of fine-tuned LLM for phase diagram generation: aLLoyM. Q\&As were generated from CPDDB using CALPHAD assessments, and Mistral was fine-tuned on these pairs.}
    \label{fig:aLLoyM}
\end{figure}

In this study, we introduce aLLoyM, an LLM fine-tuned for phase diagram generation (Fig.~\ref{fig:aLLoyM}).
Our approach leverages the Computational Phase Diagram Database (CPDDB)\cite{CPDDB}, a comprehensive open-source repository published by the National Institute for Materials Science (NIMS), as the primary training corpus.
From the CPDDB, thermodynamic database (TDB) files for 389 binary and 38 ternary phase diagrams were obtained, with the distribution of constituent elements illustrated in Figs.~S1 and S2.
Each TDB file contains Gibbs free energy functions for individual phases, enabling the construction of phase diagrams through CALPHAD assessments. 
Phase diagram calculations were performed across systematic compositional and temperature grids using Pandat software\cite{Pandat}. 
For compositional variables, elemental fractions were sampled from 0\% to 100\% in 2\% increments. 
For binaries, temperature was varied from 200 K to 5000 K in 50 K intervals, while for ternaries, the temperature was fixed at 800 K.
This systematic sampling approach generated 837,475 data points, each defining the relationship between elemental composition, temperature, and corresponding phase names.
From these data points, we constructed question–and-answer (Q\&A) pairs. For example, a question might include the information about the composition and temperature, and the answer would be the associated phase name. 
One of the important features of LLMs is their ability to handle multiple tasks within a single model. Thus, in this study, we developed a model capable of performing three different Q$\&$A tasks using a unified architecture.
We then fine-tuned Mistral, an open-source pre-trained LLM, on these Q\&As to incorporate domain-specific knowledge through selective parameter updates.

The aLLoyM model was comprehensively benchmarked using two distinct Q\&A formats: multiple-choice and short-answer. The multiple-choice Q\&As facilitated direct comparative analysis between baseline and fine-tuned model performance, with results demonstrating that fine-tuning yielded substantial improvements in predictive accuracy relative to the baseline LLM. In contrast, the short-answer Q\&As operates independently of multiple-choice constraints, rendering it particularly suitable for predicting previously unexplored phase diagrams without requiring additional domain knowledge. The implementation of the short-answer Q\&As with aLLoyM enabled the prediction of novel phase diagrams, as exemplified in Fig.~\ref{fig:aLLoyM}, and facilitated the generation of  illustrative examples. The aLLoyM model optimized for short-answer applications is publicly accessible through the Hugging Face platform (https://huggingface.co/Playingyoyo/aLLoyM).

\newpage

\section*{Results}

\subsection*{Multiple choice Q\&As}

\begin{figure}[H]
    \centering \includegraphics[width=0.7\linewidth]{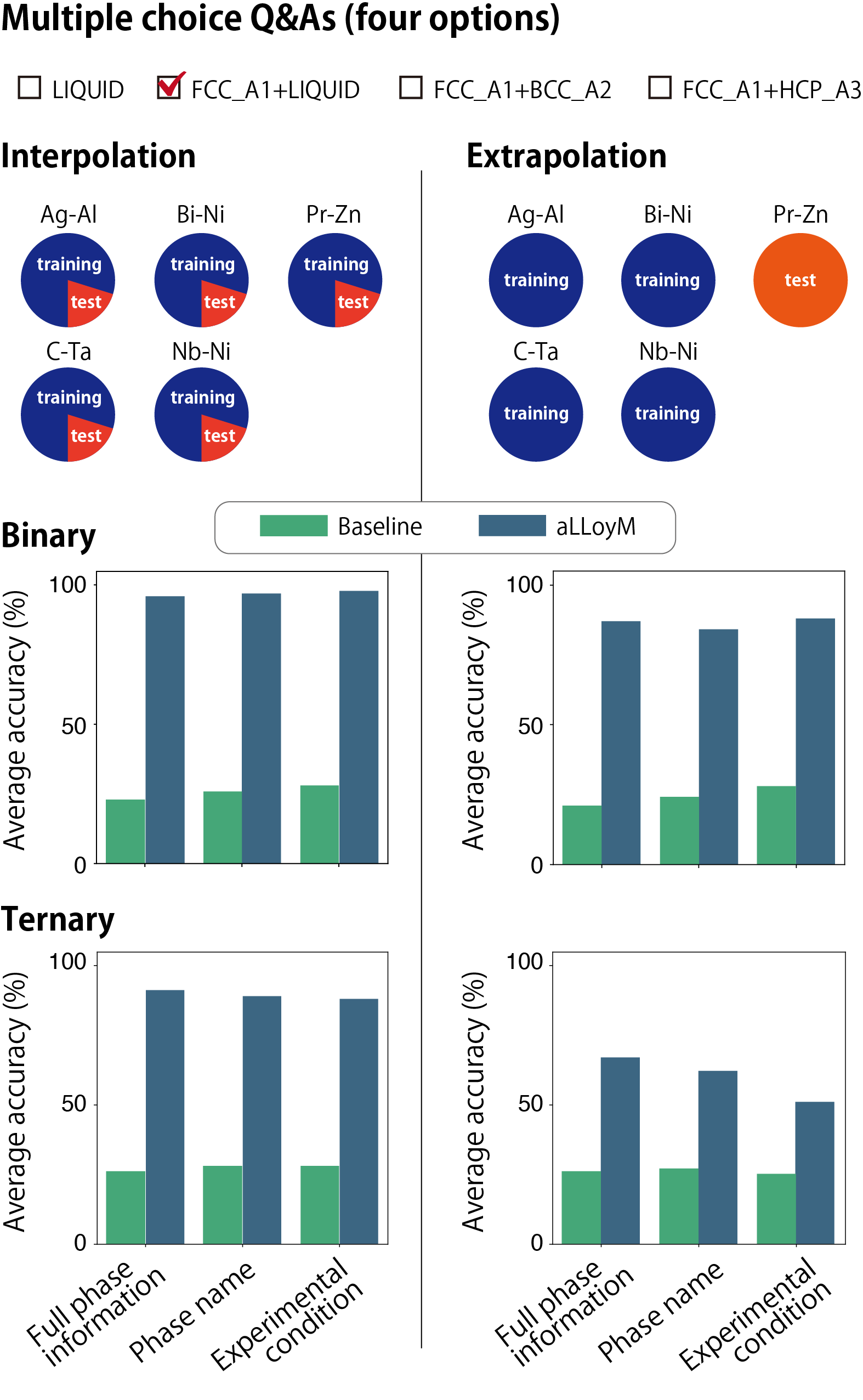}
    \caption{Accuracies of the baseline model (Mistral) and the fine-tuned model (aLLoyM) on multiple-choice Q$\&$As. Results are reported separately for interpolation and extrapolation settings, and cover all three Q$\&$A task types: full phase information inference, phase name prediction, and experimental condition inference. Performance is also distinguished between binary and ternary systems.}
    \label{fig:multi}
\end{figure}

We conducted a benchmark evaluation using multiple-choice questions to compare the performance of aLLoyM against a baseline LLM. Each question required the model to choose the correct answer from four options, where three distractors were randomly selected from answers related to the same systems (Fig.~\ref{fig:multi}). To assess the model's generalization capability, the dataset, comprising binary and ternary systems, was split into training and test sets using an 8:2 ratio. Two distinct data splitting strategies were implemented to evaluate model performance under different generalization scenarios (see Fig.~\ref{fig:multi}). 
{\it Interpolation split}: Data points were randomly distributed across all available systems, allowing assessment of model performance on familiar systems with varying compositional and thermal conditions.
{\it Extrapolation split}: Systems in the test set were completely excluded from the training set, enabling evaluation of the model's ability to generalize to previously unseen systems.

We considered three types of Q\&A tasks. 
{\it Full phase information}: Given the input composition and temperature, the model predicts the complete phase information, including phase names and their corresponding fractions and compositions. An example of this Q\&A task is presented in Fig.~\ref{fig:aLLoyM}.
{\it Phase name}: The model predicts only the phase names based on the input composition and temperature. The output is the phase domain without specifying phase fractions or compositions for full phase information.
{\it Experimental condition}: Given the constitutive elements and a specific phase domain, the model predicts a possible composition and temperature. This task serves as the inverse of the phase name prediction.
The examples of each Q$\&$A are summarized in Table~S1. These three Q$\&$A tasks were trained within a single LLM.

The accuracies of all Q\&A tasks in the multiple-choice are shown in Fig.~\ref{fig:multi}, with performance evaluated separately for binary and ternary systems across the three task types. As a baseline, we employed the Mistral-Nemo-Instruct-2407-bnb-4bit model using Hugging Face’s causal language modeling interface\cite{mistral-nemo}. The baseline model's performance remained close to random guessing in both interpolation and extrapolation settings, with accuracy only slightly above the level expected by chance. These findings indicate that the baseline language model struggled to produce correct answers to phase diagram questions.

In contrast to the baseline, the fine-tuned models exhibited substantial performance improvements across all tasks. For both interpolation and extrapolation settings, individual models were fine-tuned on the complete ensemble of three Q\&A tasks. In all cases, the fine-tuned models outperformed the baseline. As anticipated, performance was generally higher on interpolation tasks compared to extrapolation tasks. Furthermore, predictions for ternary systems proved more challenging than those for binary systems, while performance differences among the three Q\&A tasks were relatively minor. These results demonstrate that, when provided with suitable training data, LLMs are capable of accurately predicting phase diagram information. Notably, the model’s success on extrapolation setting suggests an ability to generalize knowledge from known systems to make informed predictions for previously unseen combinations.

\subsection*{Short answer Q\&As}

Adopting the short-answer questions allows the model to generate responses without relying on predefined multiple-choice options. Consistent with the multiple-choice Q\&As, the fine-tuned models were trained using the full data corresponding to all three Q\&A tasks. To evaluate the alignment between the ground-truth answers and those generated by aLLoyM, we introduced a scoring metric described in the Methods section depending on the Q\&A task. The score ranges from 0\% to 100\%, with higher values indicating greater agreement between the generated and ground-truth answers.

\begin{figure}[H]
    \centering \includegraphics[width=0.7\linewidth]{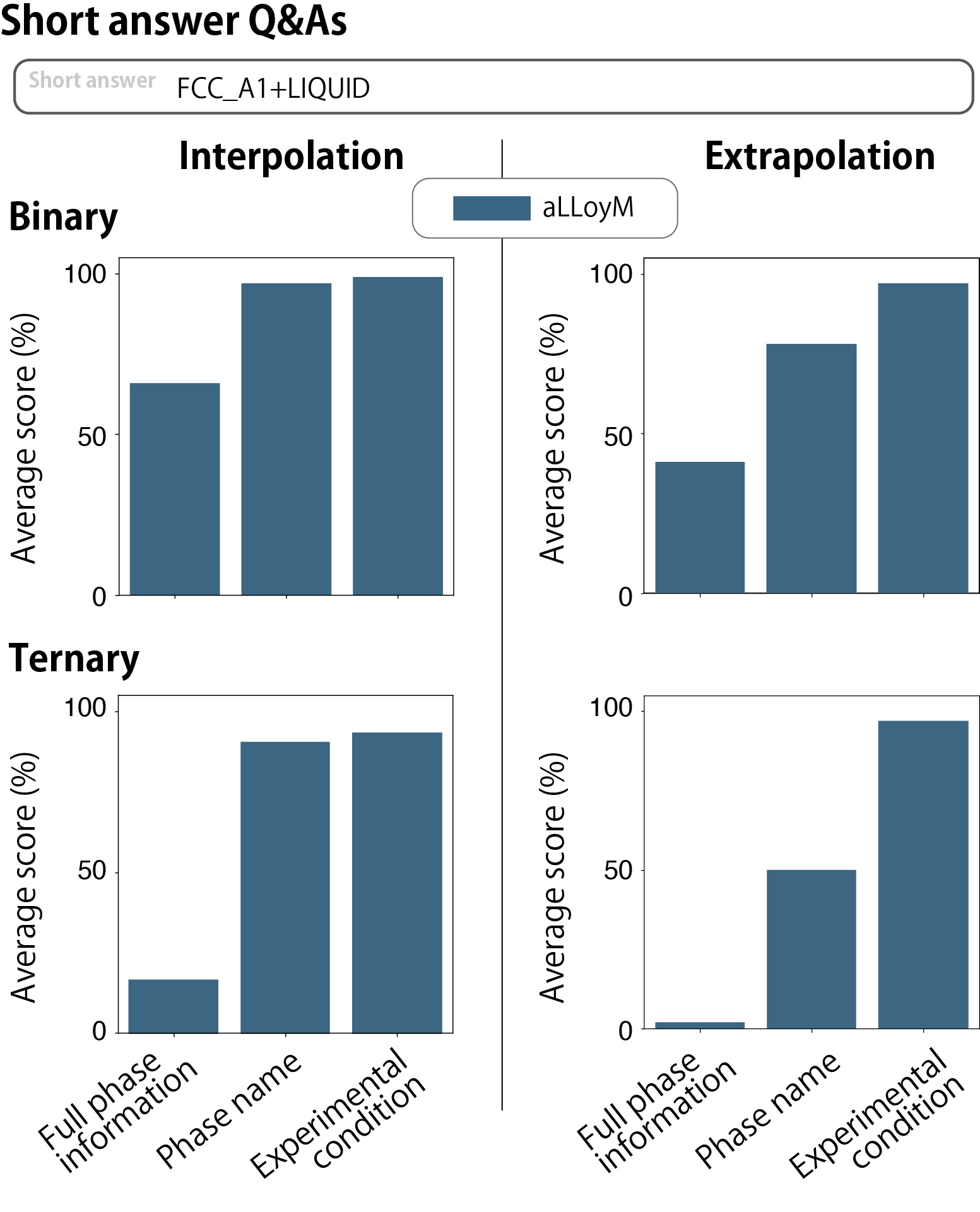}
    \caption{Average scores of the fine-tuned models (aLLoyM) for short answer Q\&As.
    Results are presented for interpolation and extrapolation configurations, with individual tasks (full phase information, phase name, and experimental condition).
    Binary and ternary systems were evaluated separately.}
    \label{fig:short}
\end{figure}

Figure~\ref{fig:short} presents the average scores for each task. As anticipated, performance was superior on interpolation settings relative to extrapolation settings. Among the three Q\&A task categories, predicting complete phase information proved most challenging. Nevertheless, the model demonstrated robust performance in predicting phase names, even under extrapolation conditions. Furthermore, it successfully generated appropriate experimental conditions from specified phase information in extrapolation settings, suggesting that when a target phase is designated, aLLoyM possesses the capacity to reliably propose suitable experimental parameters. Across all tasks, predictions for ternary systems were consistently more challenging than those for binary systems.

\begin{figure}
    \centering
    \includegraphics[width=1\linewidth]{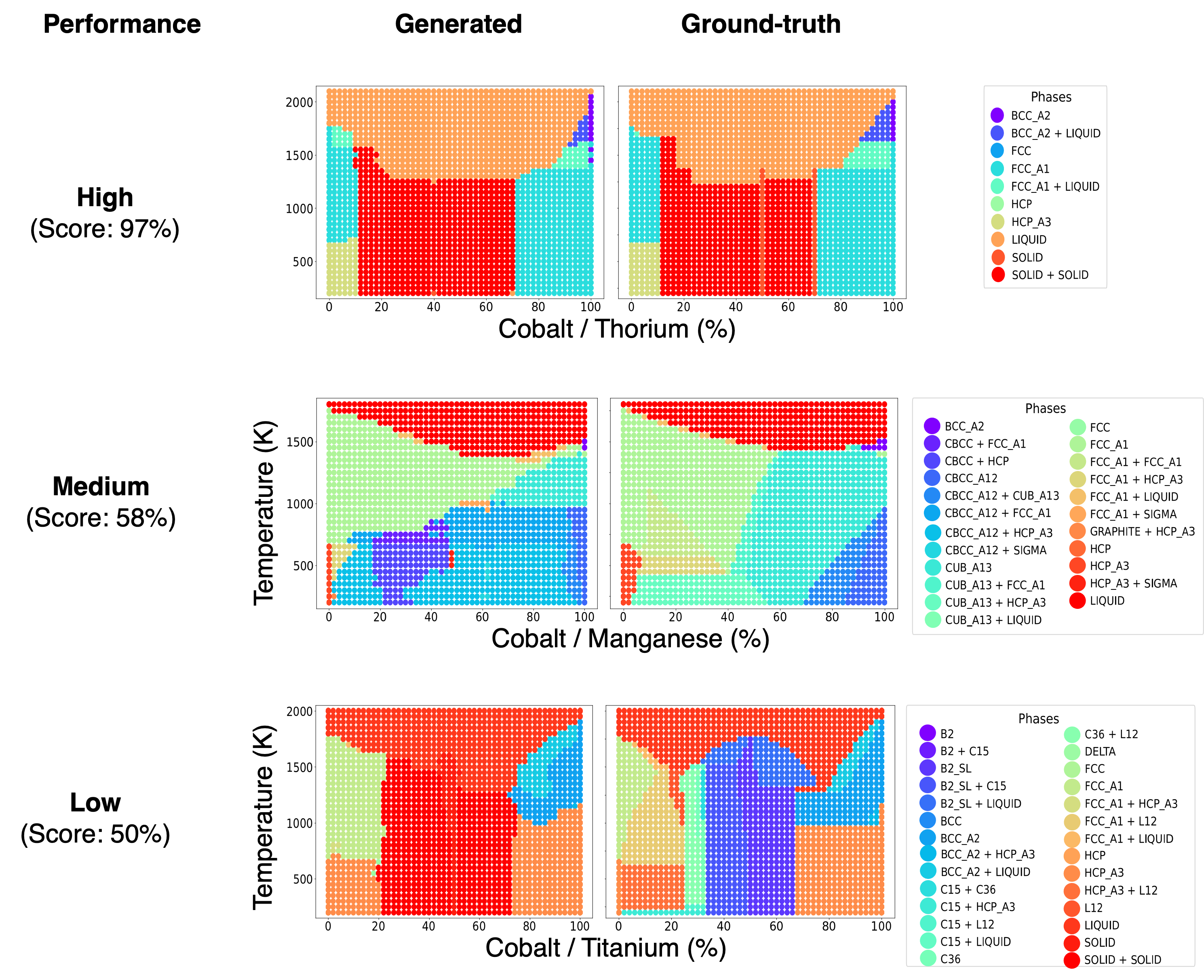}
    \caption{Representative binary phase diagrams exhibiting varying predictive performance, as generated by aLLoyM for the phase name prediction task. The ground-truth phase diagrams are also shown. Lower scores denote greater discrepancies between generated and ground-truth phase diagrams.}
    \label{fig:binary}
\end{figure}

\begin{figure}
    \centering
    \includegraphics[width=1\linewidth]{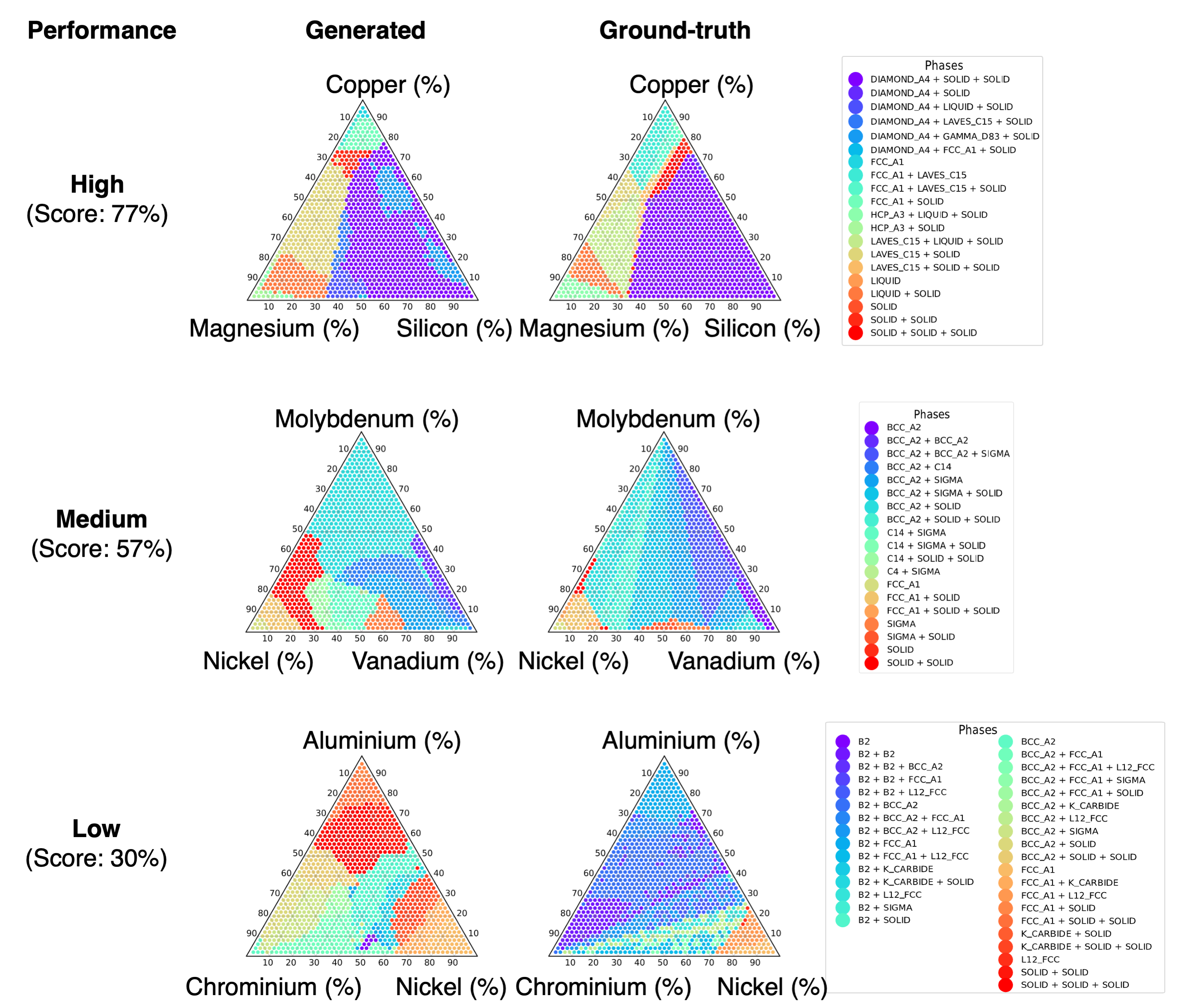}
    \caption{Representative 800K ternary isothermal sections exhibiting varying predictive performance, as generated by aLLoyM for the phase name prediction task. The ground-truth phase diagrams are also shown. Lower scores denote greater discrepancies between generated and ground-truth phase diagrams.}
    \label{fig:ternary}
\end{figure}

Based on the phase names predicted by aLLoyM, we reconstructed the phase diagrams for the element sets in the extrapolated test set. Figures ~\ref{fig:binary} and ~\ref{fig:ternary} present representative binary and ternary phase diagrams exhibiting varying levels of predictive performance. The scores represent averages across each complete phase diagram, with corresponding ground-truth phase diagrams provided for comparison.
Across all cases, predictive performance remains consistently higher in regions proximate to pure elements and diminishes progressively as compositions approach intermediate regions. When the intermediate compositional range exhibits relatively simple phase behavior, the generated phase diagrams demonstrate greater accuracy, yielding elevated scores as observed in the Co-Th and Mg-Si-Cu systems. Conversely, systems characterized by more complex intermediate phase behavior frequently produce lower scores, as exemplified by the Co-Ti and Cr-Ni-Al systems. These findings indicate that the inherent complexity of intermediate compositional regions is a key factor contributing to the difficulty of phase diagram generation for aLLoyM.

\begin{figure}[H]
    \centering
    \includegraphics[width=1\linewidth]{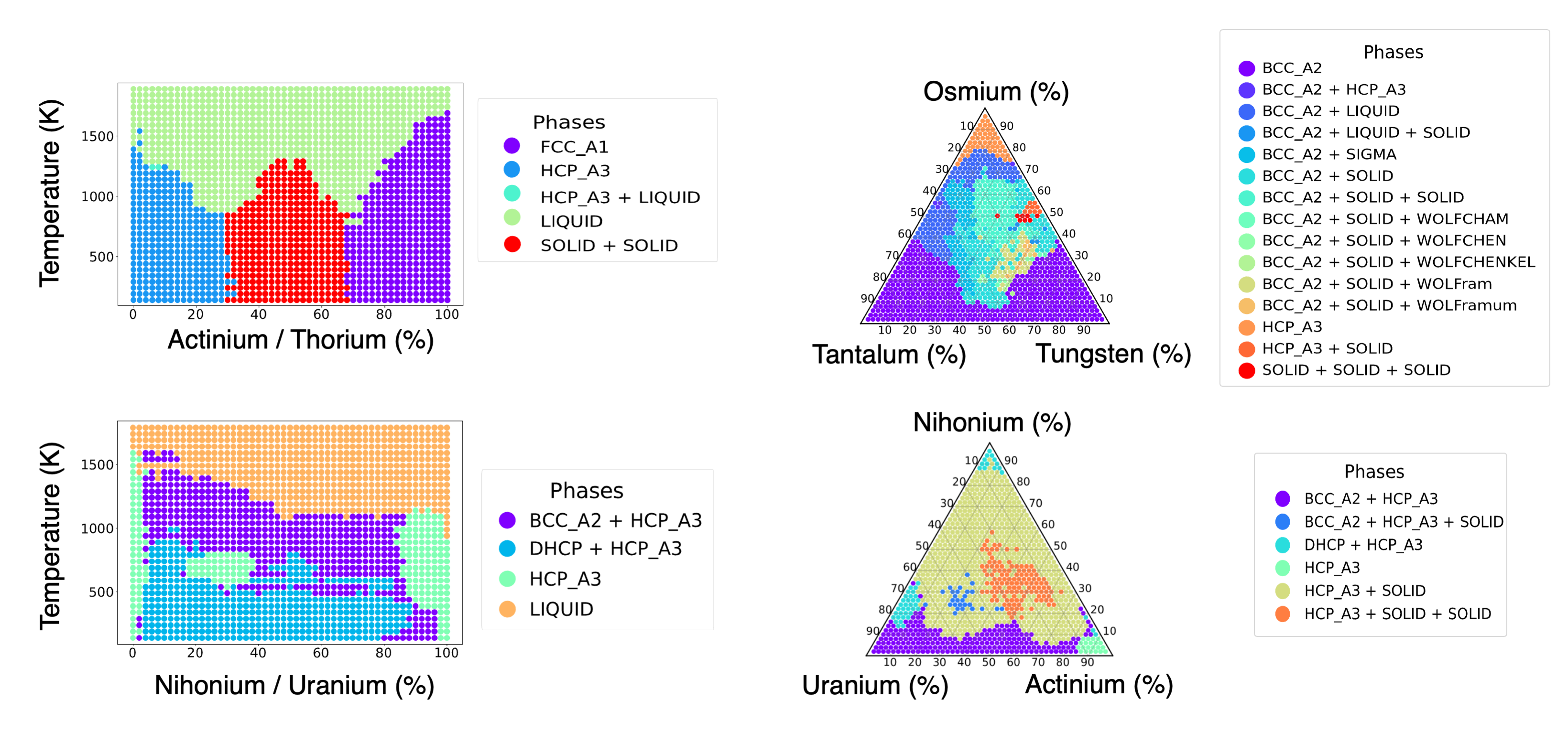}
    \caption{Examples of unknown binary phase diagrams and 800K ternary isothermal sections inferred by aLLoyM in the phase name prediction task.}
    \label{fig:phase_diagram_unknown}
\end{figure}

\subsection*{Novel phase diagram generation}

aLLoyM enables the generation of entirely novel phase diagrams, including those that are currently unknown or extremely difficult to construct experimentally. Figure~\ref{fig:phase_diagram_unknown} presents examples of such phase diagrams for both binary and ternary systems, generated using aLLoyM with the short-answer Q\&A format.
We first examine the results for binary systems. Phase diagrams were generated for the Th–Ac (thorium–actinium) and U–Nh (uranium–nihonium) systems. In the case of Th–Ac, pure thorium was incorporated within the training dataset, whereas actinium was omitted owing to its short half-life. aLLoyM predicted the melting point of actinium to be approximately 1400$^\circ$C, which is consistent with the experimental value of approximately 1050$^\circ$C\cite{actinium}. However, while the stable crystal structure of actinium is known to be face-centered cubic (FCC)\cite{Farr1961}, the model incorrectly predicted it as hexagonal close-packed (HCP).
For the U–Nh system, neither uranium nor nihonium was included in the training data, making this an entirely extrapolative prediction. The predicted melting point for uranium was approximately 900$^\circ$C, compared to the known value of 1135$^\circ$C\cite{Uranium}, indicating only a moderate deviation. However, aLLoyM erroneously predicted HCP as the stable structure, while uranium's actual stable structure is body-centered cubic (BCC)\cite{Grenthe}. For nihonium, no experimental data on melting point or crystal structure are currently available. Nevertheless, the model was able to generate phase diagram outputs, illustrating its potential to make predictions in domains where experimental data are scarce or nonexistent. This highlights the promise of LLMs as tools for exploratory materials design and discovery.

We subsequently examine the results for ternary systems. Tungsten (W), tantalum (Ta), and osmium (Os) are all elements characterized by exceptionally high melting points, rendering experimental investigation of their ternary phase diagram particularly challenging. To date, no ternary phase diagrams have been established for this system, although all three constituent elements are present in the training data for binary systems. Using aLLoyM, we reconstructed the ternary phase diagram for this system at 800 K. In the intermediate compositional region, the model predicts the emergence of three-phase coexistence. Notably, aLLoyM also predicts the existence of phases designated with ``WOLF'' nomenclature that are absent from the training data. These may reflect latent knowledge embedded within the pre-trained Mistral model. Finally, we generated a ternary phase diagram for nihonium, uranium, and actinium at 800 K, representing an entirely hypothetical system that cannot be experimentally realized. Here as well, the model predicts three-phase coexistence in the intermediate compositional region as well as the W-Ta-Os system.

The phase diagrams generated above remain beyond experimental validation. Nevertheless, this transformative technology, which can estimate novel phase diagrams solely based on constituent elements, has the potential to become a valuable tool for rational materials design.

\clearpage

\section*{Discussion}

In this work, we developed aLLoyM, a fine-tuned Large Language Model specialized for relations between alloy compositions, temperatures, and phase information. The model was fine-tuned on Q\&As for binary and ternary phase diagrams constructed from the open-source Computational Phase Diagram Database (CPDDB) using CALculation of PHAse Diagrams (CALPHAD) assessments. Our benchmark results demonstrated that fine-tuning significantly improves the accuracy of the model in selecting the correct responses to multiple-choice questions concerning phase diagrams. Furthermore, the short-answer model of aLLoyM demonstrated the capacity to generate phase diagrams for previously unreported systems. These findings establish aLLoyM as a promising foundation for phase diagram prediction, with demonstrated capabilities for extrapolation to novel systems. The capability to infer phase behavior in unexplored compositional spaces has the potential to significantly accelerate the design and discovery of new materials.

The consistently stronger performance on binary systems compared to ternary systems across all evaluations can be attributed to the relatively limited availability of ternary training data. Future work should therefore prioritize the expansion of training resources for ternary and higher-order systems to enable robust multi-component phase diagram prediction.
A key advantage of aLLoyM’s natural language framework lies in its ability to utilize the virtually unlimited vocabulary of elements and phase names acquired during pretraining, thus supporting broad generalization to diverse chemical systems. While the current implementation tends to generate phase names seen during training, this limitation opens promising opportunities for improvement through advanced prompt engineering. In particular, incorporating thermodynamics-aware prompts may help guide the model toward applying physically meaningful reasoning during inference, thereby enhancing prediction accuracy.
Collectively, these directions represent important pathways for future research toward developing more capable LLMs specifically tailored to phase diagram prediction and materials discovery.

\clearpage

\section*{Methods}

\subsection*{Fine-tuning}

We fine-tuned the Mistral-Nemo-Instruct-2407 model using LoRA (Low-Rank Adaptation) with rank 16 and alpha 16, targeting attention and feed-forward projections. Training data was formatted using a structured prompt template with Instruction, Input, and Output sections (see Table S1). The model was trained for 15,000 steps with a learning rate of \(2 \times 10^{-4}\), batch size of 16 per device, and 4 gradient accumulation steps using the AdamW optimizer with bfloat16 precision.

\subsection*{Scoring criteria for generated answers}

The scoring criteria for generated answers depends on the specific Q$\&$A task, as both the answer format and the target subject vary across tasks. The definition of the scores for each task is shown below.

\textbf{Full phase information:} This requires an exact match between the generated answer and the ground-truth answer. The score is defined as
\[
\text{Score} = 
\begin{cases}
100\% & \text{: if the generated answer exactly matches the ground truth} \\
0\%& \text{: otherwise}
\end{cases}.
\]

\textbf{Phase name:} We first construct lists of all phase domains in the generated answer and the ground truth, denoted as $A$ and $B$, respectively.
Let $k$ and $l$ be the number of phase domains in the generated answer and the ground truth, respectively.
The resulting lists are as follows:
\begin{align}
A &= [\text{phase name} \ 1, \ \text{phase name} \ 2, \ ..., \ \text{phase name} \ k], \notag \\
B &= [\text{phase name} \ 1, \ \text{phase name} \ 2, \ ..., \ \text{phase name} \ l]. \notag
\end{align}
The score is then evaluated using the Jaccard similarity of perfectly matching phase names, defined as:
\[
\text{Score} = 
\begin{cases}
|A \cap B| / |A \cup B| \times 100\% & \text{: if } |A \cup B| > 0 \\
0\% & \text{: otherwise}
\end{cases}.
\]
With this definition, scores can be evaluated even when the number of phase domains differs between the generated answer and the ground truth.

\textbf{Experimental condition:} The scoring of experimental conditions evaluates how well the element compositions and temperature match one of the ground-truths by comparing composition accuracy and temperature accuracy.
\textit{Composition Accuracy} measures how closely the generated composition matches the ground-truth composition. 
The compositions of the generated answer and the ground-truth are defined as $\{ x_n \}_{n=1,...,N}$ and $\{ x^*_n \}_{n=1,...,N}$ where $N$ is the number of elements in the target system.
The composition accuracy is defined as
\[
A_{\rm c} = \frac{1}{N} \sum_{n = 1}^N \left( 1 - \frac{| x_n^* - x_n |}{100} \right).
\]
To standardize the values, the each composition difference $| x_n^* - x_n |$ is divided by 100 which is the maximum difference.
\textit{Temperature Accuracy} measures how closely the generated temperature $T$ matches the ground-truth temperature $T^*$. It normalizes the absolute difference $\Delta T$ by the full range of temperatures available in the target elemental system.
The temperature accuracy is defined as
\[
A_{\rm t} =  1 - \frac{| T^* - T |}{\Delta T}.
\]
Using these accuracies, the score for experimental condition is defined as
\[
\text{Score} = \left( \frac{1}{2} A_{\rm c} + \frac{1}{2} A_{\rm t} \right) \times 100\%.
\]
This score is 100\% when both the compositions and the temperature in the generated answer exactly match with any of the ground truths.
Note that in this case, multiple ground-truth answers may exist.
Therefore, the above score is computed for each ground-truth answer, and the maximum value is adopted.

\subsection*{Data availability}
The data used in this paper, as well as the short-answer
fine-tuned version of aLLoyM are available at \url{https://huggingface.co/Playingyoyo/aLLoyM}.

\subsection*{Code availability}
Code for aLLoyM is available at
\url{https://github.com/tsudalab/aLLoyM/tree/main}.

%Reference

\subsection*{Acknowledgements}
The authors would like to thank Etsuko Ogamino for data collection.
This study was supported by a project subsidized by JSPS KAKENHI (25K01492 and 25KJ0870), JST-CREST (JPMJCR21O2), and MEXT Program: Data Creation and Utilization Type Material Research and Development Project.

\subsection*{Author contributions}
All the authors conceived the original idea.
Y.O. prepared Q$\&$As from CALPHAD assessment data and developed aLLoyM.
G.D., T.A., and R.T. prepared the CALPHAD assessment data of phase diagrams.
Y.O, R.T., and K.T. wrote the original manuscript. All the authors discussed the results, commented on the manuscript, and approved the final version of the manuscript.

\newpage

\renewcommand{\thefigure}{S\arabic{figure}}
\renewcommand{\thetable}{S\arabic{table}}

\setcounter{figure}{0}
\setcounter{table}{0}

\section*{Supplementary information}

\begin{figure}[H]
\centering
\includegraphics[width=0.8\linewidth]{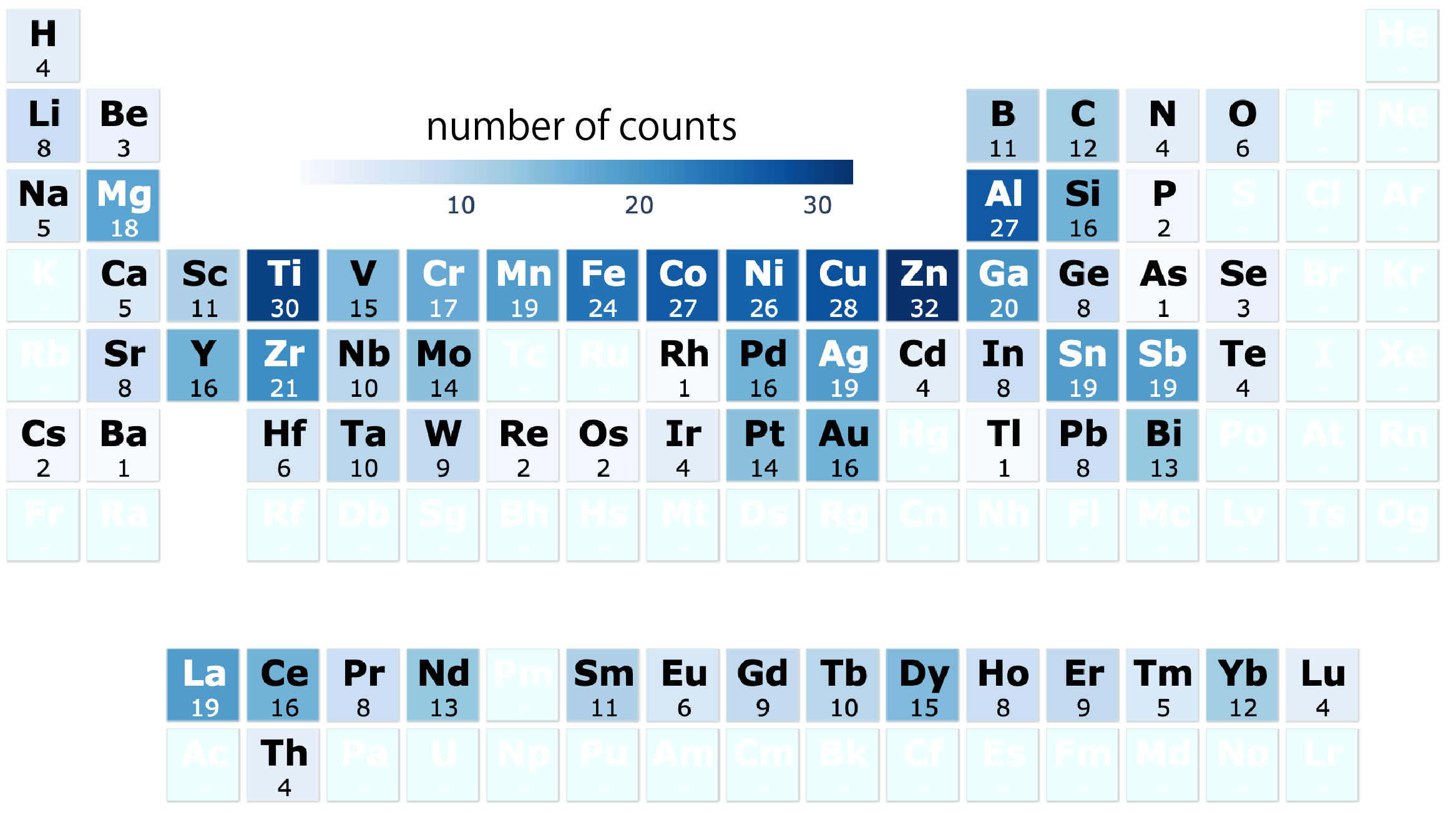}
\caption{Distribution of elements in the binary dataset. A total of 389 binary phase diagrams were extracted from CPDDB.}
\label{fig:periodic}
\end{figure}

\begin{figure}[H]
\centering
\includegraphics[width=0.8\linewidth]{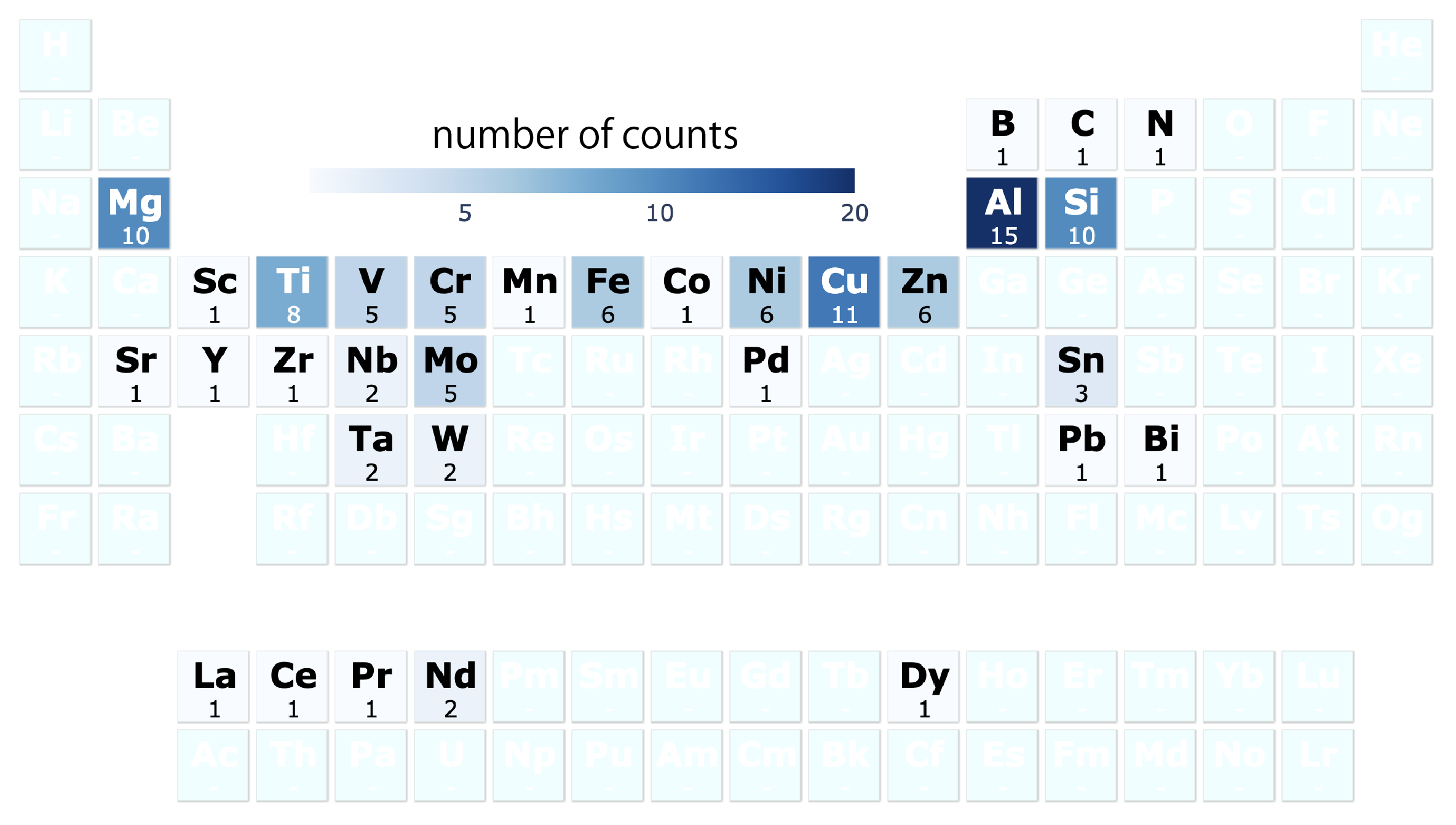}
\caption{Distribution of elements in the ternary dataset. A total of 38 ternary phase diagrams were extracted from CPDDB.}
\label{fig:periodic}
\end{figure}

\begin{table}[H]
\centering
\begin{tabular}[H]{|>{\columncolor{gray!20}}m{2.0cm}|p{4.2cm}|p{4.2cm}|p{4.2cm}|}
\hline
Task & Full phase information & Phase name & Experimental condition \\
\hline
Instruction & \multicolumn{3}{p{12.9cm}|}
{You are an expert in phase diagrams, thermodynamics, and materials science, specializing in alloy systems.} \\
\hline
Question (Input) & What phases form when Silver (34\%) + Yttrium (66\%) are mixed at 500 K? 
& What phases form when Silver (34\%) + Yttrium (66\%) are mixed at 500 K? Answer phase names only. 
& Under what condition do Silver + Yttrium form HCP\_A3 + SOLID? \\
\hline
Answer (Output) & 32\% HCP\_A3 + 68\% SOLID with composition ratio Silver : Yttrium = 1 : 1.
& HCP\_A3 + SOLID.
& Silver (34\%) + Yttrium (66\%) mixed at 500 K. \\
\hline
\end{tabular}
\caption{Excerpts of three types of Q\&As.
Instruction is common through three tasks.}
\label{tab:short_answer_examples}
\end{table}

\end{document}